\documentstyle[11pt]{article}
\begin{document}
 \newcommand{\be}{\begin{equation}} \newcommand{\fe}{\end{equation}}
\newcommand{\eqn}{\label}\newcommand{\bel}{\begin{equation}\label}

%\def\thf{\baselineskip=\normalbaselineskip\multiply\baselineskip
%by 7\divide\baselineskip by 6}
%\thf
 
%\lta and \gta produce > and < signs with twiddle underneath
\def\spose#1{\hbox to 0pt{#1\hss}}\def\lta{\mathrel{\spose{\lower 3pt\hbox
{$\mathchar"218$}}\raise 2.0pt\hbox{$\mathchar"13C$}}}  \def\gta{\mathrel
{\spose{\lower 3pt\hbox{$\mathchar"218$}}\raise 2.0pt\hbox{$\mathchar"13E$}}}

\noindent

{\bf RECENT DEVELOPMENTS IN VORTON THEORY}

\medskip
{\bf Brandon Carter}

\medskip
{\bf D.A.R.C., Observatoire de Paris}

{\bf 92 Meudon, France}

\medskip
{ Contribution to 1996 Peyresq Meeting}

\bigskip {\bf Astract} {\it This article provides a concise overview of
recent theoretical results concerning the theory of vortons, which are
defined to be (centrifugally supported) equilibrium configurations of
(current carrying) cosmic string loops. Following a presentation of the
results of work on the dynamical evolution of small circular string
loops, whose minimum energy states are the simplest examples of
vortons, recent order of magnitude estimates of the cosmological
density of vortons produced in various kinds of theoretical scenario
are briefly summarised.}

\bigskip \parindent =2 cm

\section{introduction}

It is rather generally accepted\cite{book} that among the conceivable
varieties of topological defects of the vacuum that might have been generated
in early phase transitions, the {\it vortex type} defects describable on a
macrosopic scale as {\it cosmic strings} are the kind that is most likely to
actually occur -- at least in the post inflationary epoch -- because the other
main categories, namely monopoles and walls, would produce a catastrophic
cosmological mass excess. Even a single wall stretching accross a Hubble
radius would by itself be too much, while in the case of monopoles it is their
collective density that would be too high unless the relevant phase transition
occurred at an energy far below that of the G.U.T. level, a possibility that
is commonly neglected on the grounds that no monopole formation occurs in the
usual models for the transitions in the relevant range, of which the most
important is that of electroweak symmetry breaking. 

The case of cosmic strings is different. One reason is that (although
they are not produced in the standard electroweak model) strings are
actually produced at the electroweak level in many of the commonly
considered (e.g.  supersymmetric) alternative models. A more commonly
quoted reason why the case of strings should be different, even if they
were formed at the G.U.T level, is that -- while it may have an
important effect in the short run as a seed for galaxy formation --
such a string cannot be cosmologically dangerous just by itself, while
a distribution of cosmic strings is also cosmologically harmless
because (unlike ``local'' as opposed to ``global'' monopoles) they will
ultimately radiate away  their energy and progressively disappear.
However while this latter consideration is indeed valid in the case of
ordinary Goto-Nambu type strings, it was  pointed out by Davis and
Shellard\cite{DS} that it need not apply to ``superconducting''
current-carrying strings of the kind originally introduced by
Witten\cite{witten}. This is because the occurrence of stable currents
allows loops of string to be stabilized in states known as ``vortons'',
so that they cease to radiate.

The way this happens  is that the current, whether timelike or
spacelike, breaks the Lorentz invariance along the string worldsheet
\cite{for,mal,neutral,enon0}, thereby leading to the possibility of
rotation, with velocity $v$ say. The centrifugal effect of this
rotation, may then compensate the string tension $T$ in such a way as
to produce an equilibrium configuration, i.e. what is known as a {\it
vorton}, in which \be T= v^2 U \, , \eqn{1}\fe 
where $U$ is the energy per unit length in the corotating rest
frame\cite{mal,ring}. This condition is interpretable as meaning that
the circulation velocity $v$ is the same as the velocity\cite{for} of
extrinsic (transverse) ``wiggle'' type perturbation -- so that relatively
backward moving ``wiggles'' will be effectively static deformations.
Such a vorton state will be stable, at least classically, if it
minimises the energy for given values of the pair of conserved
quantities characterising the current in the loop, namely the phase
winding number $N$ say, and the corresponding particle number $Z$ say,
whose product

\be J=NZ \eqn{2}\fe
is interpretable in the case of a circular loop as the magnitude of its
angular momentum. 
 If the current is electromagnetically
coupled, with charge coupling constant $e$, then the latter will determine a 
corresponding vorton charge $Q=Ze$.

Whereas the collective energy density of a distribution of
non-conducting cosmic strings will decay in a similar manner to that of
a radiation gas, in contrast, for a  distribution of relic vortons, the
energy density will scale like that of ordinary matter. Thus, depending
on when and how efficiently they were formed, and on how stable they
are in the long run, such a distribution of vortons might eventually
come to dominate the density of the universe.  It has been rigorously
established\cite{stabgen,stab,stabwit} that circular vorton
configurations of this kind will commonly (though not always) be stable
in the dynamic sense at the classical level, but very little is known
so far about non-circular configurations or about the question of
stability against quantum tunnelling effects, one of the difficulties
being that the latter is likely to be sensitively model dependent.

\section{Dynamics of circular string loops.}

Using the kind of string model\cite{CP95} that is appropriate for
describing the effect of a Witten type current, Larsen and
Axenides\cite{LA96} have recently carried out an analytic treatment of
the special case of a free circular string loop for which the angular
momentum $J$ vanishes, so that there is no centrifugal effect to
prevent the ultimate collapse of the loop.  Extending this to the case
for which  a centrifugal is present barrier, Peter, Gangui, and the
present author  have undertaken a systematic analysis\cite{CPG96} of the
dynamics of a free circular string loop carrying a Witten type in the
generic case for which both $N$ and $Z$ have non-zero values.  These
values will be given in terms of a pair of Bernoulli type constants of
integration, $B$ and $C$ say, by
\be N={B\over 2\pi\sqrt{\kappa_{_0}}}\, ,\hskip 1 cm Z= C\sqrt{\kappa_{_0}}
\, ,\eqn{3}\fe
where $\kappa_{_0}$ is a constant of the order of unity that depends on
the particular form of the string model.  The kind of string model
needed for representing the macroscopic effect of a Witten type
``superconducting'' vacuum vortex is specified\cite{CP95} by a
Lagrangian function ${\cal L}$ depending on a single scalar variable
$w$ that is proportional to the squared magnitude of the the scalar
phase variable (whose loop integral specifies the winding number $N$)
with $\kappa_{_0}$ as the proportionality constant. An important role
in the analysis is played by derived function ${\cal K}$ that is
obtainable the Lagrangian using the formula
\be {\cal K}=-\Big(2{d{\cal L}\over dw}\Big)^{-1} \, ,\eqn{4}\fe
and that is adjusted (by the convention used to fix the normalisation
of $\kappa_{_0}$) so that ${\cal K}$ tends to unity in the null current
limit, i.e. as $w$ tends to zero.

The solutions fall generically into two distinct classes, namely those
for which $B^2>C^2$, in which the string current remains always
spacelike, i.e. $w>0$, and those for which $B^2<C^2$, in which the
current remains always timelike, i.e. $w<0$.  Intermediate between
these classes is the special ``chiral'' limit case characterised by
$B^2=C^2$, in which the current remains always null, i.e.  $w=0$.  A
key intermediate step in the analysis is the establishment of a
relation of the form
\be \ell^2 ={B^2 -C^2 {\cal K}^2\over w}\, ,\eqn{5}\fe
which determines the loop circumference $\ell=2\pi r$ as a function of $w$ and 
thus implicitly determines the state variable $w$ as a function of $\ell$,
except in the special ``chiral''  limit case $B^2=C^2$ for which $w=0$
so that the right hand side is indeterminate.
The outcome of the analysis\cite{CPG96} is that
that the string radius $r$ is governed by an equation of the form
\be M^2\dot r^2= M^2 - \Upsilon^2\, ,\eqn{6}\fe
(using a dot for time differentiation) in which $M$ is the constant mass energy
of the loop and $\Upsilon$ is an effective potential that is determined 
as a function of the loop circumference in the form
\be \Upsilon= {C^2{\cal K}\over\ell} - {\cal L}\ell\, ,\eqn{7}\fe
where the coefficients ${\cal K}$ and $\cal L$ have constant values
(the former being just unity) in the ``chiral'' case, while in general
they will be determined as (rather slowly varying) functions of $\ell$
by the relation (\ref{5}). It is to be observed that there is an
asymptotic confining potential contribution that rises linearly with
the radius, and that (unlike the familiar Kepplerian particle problem
for whch the centrifugal barrier goes as the inverse square of the
radius) in the present case the effectivecentrifugal barrier goes just
as the inverse first power of the radius. 

It is evident in the ``chiral'' case, and can be verified in general,
that the effective potential given by (\ref{7}) will have a gradient given by
\be    {d\Upsilon\over d\ell}= -{{\cal B}^2\over{\cal K}\ell^2}-{\cal L}
\, ,\eqn{8}\fe 
which vanishes at the minimum of $\Upsilon$ where
\be B^2\Big({\cal L}+{w\over{\cal K}}\Big)=C^2{\cal K}^2{\cal L}\, .\eqn{9}\fe
It can be seen from (\ref{6}) that the value of $\Upsilon$ at this
minimum will also be the minimum admissible value of the mass parameter for
the given values of $B$ and $C$, and that when $M$ has this minimum
value the loop will be in a state of equilibrium with radius given by
the solution of (\ref{9}), which can be shown to be equivalent to the
vorton equilibrium condition (\ref{1}) that was quoted at the outset.

Whether such a vorton equilibrium state is actally attainable depends on
whether the minimality condition (\ref{9}) actually has a solution in
the admissible range of string states. The explicit form of the relevant
string model\cite{CP95} is given by 
\be {\cal L}= -m^2 - m_\ast^{\, 2}\ln\sqrt{\cal K}\, ,\eqn{10}\fe
with
\be {\cal K}=1+{w\over m_\ast^{\, 2}}\, ,\eqn{11}\fe
where $m$ and $m_\ast$ are constants. The first one, $m$, is the
relevant Kibble mass, as characterised by the condition that $m^2$
should be the common limit of the tension $T$ and the energy density
$U$ in the zero current limit. The other one, $m_\ast$, is the mass
scale associated with the carrier field responsible for the Witten
current. The validity of this model is limited to the range
\be {\rm e}^{-2m^2/m_\ast^{\, 2}}<{\cal K}<2 \, ,\fe
the lower bound being where the tension $T$ tends to zero in the
timelike current regime while the upper bound is the saturation limit
in the timelike current regime. It can be seen that the loop will
oscillate periodically without approaching either of these
``dangerous'' limits provided the ratio $B/C$ is not too far from unity
(i.e. from the ``chiral''
 value).

\section{Order of magnitude estimates}

Whether or not it is exactly circular (as assumed in the analytic
treatment that has just been described) the numerical value of the
total mass $M$ of a vorton state characterised by the quantum numbers
$N$ and $Z$ will be given\cite{ring,vorton} in rough order of magnitude 
by a formula of the form
\be M\approx\vert NZ\vert^{1/2} m \eqn{12}\fe
where $m$ is the relevant Kibble mass, which will normally be given
approximately by the mass of the Higgs field responsible for the relevant
vacuum symmetry breaking.

In the earliest crude quantitative estimates\cite{DS,vorton} of the
likely properties of a cosmological vorton distribution produced in
this way, it was assumed not only that the current was stable
against leakage by tunnelling, but also that the Witten mass scale $m_\ast$
characterising the relevant carrier field was of the same order of
magnitude as the Kibble mass scale $m$ characterising the string
itself. The most significant development in the more detailed
investigations carried out more recently\cite{Moriand95,BCDM96} is the
extension to cases in which $m_\ast$ is considerably smaller than
$m$.  A rather extreme example that immediately comes to mind is that
for which $m$ is postulated to be at the G.U.T. level, while
$m_\ast$ is at the electroweak level, in which case it is found that
the resulting vorton density will be far too low to be cosmologically
significant.

The simplest scenarios are those for which (unlike the example just quoted)
the relation 
\be \sqrt{m_\ast}\gta m \eqn{13}\fe
is satisfied in dimensionless Planck units as a rough order of magnitude
inequality. In this case the current condensation would have ocurred during
the regime in which (as pointed out by Kibble\cite{kibble} in the early
years of cosmic string theory) the dynamics was dominated by friction damping.
Under these circumstance it is estimated\cite{Moriand95,BCDM96} that
the typical value of the quantum numbers of vortons in the resulting
population will be given very roughly by 
\be N\approx Z\approx  m^{1/2} m_\ast^{-3/4} \, ,\eqn{14} \fe 
which by (\ref{12}) implies a typical vorton mass given by 
\be M\approx\Big({m\over\sqrt{ m_\ast}}\Big)^{3/2} \, ,\eqn{15}\fe
which, in view of (\ref{13}), will never exceed the Plank mass. 
When the cosmological temperature has fallen to a value $\Theta$ say, the
estimated number density $n$ of the vortons is given as a constant fraction of
the corresponding number density  $\approx \Theta^3$ of black body photons by
the rough order of magnitude formula 
\be {n\over\Theta^3}\approx \Big( {\sqrt{m_\ast}\over m}\Big)^3
 m_\ast^{\,3} \, .\eqn{16}\fe 
It follows in this case that, in order to avoid producing a cosmological mass
excess, the value of $m_\ast$ in this formula should not exceed a limit that
works out to be of the order of $10^{-9}$, and the limit is even be smaller,
around $10^{-11}$, when the two scales $m_\ast$ and $m$ are comparable.

Limits in roughly the same range, round about $10^{-10}$ (about
midway between the G.U.T. value $10^{-3}$ and the electroweak value
$10^{-16}$) are also obtained\cite{BCDM96} for $m_\sigma$ in scenarios
for which, instead of (\ref{13}), 
\be \sqrt{m_\ast} \ll m\, , \eqn{17}\fe
but in this case the analysis is more complicated and also 
more uncertain, since the result
is sensitive to the value of an efficiency factor $\varepsilon$ 
that is expected to be of order unity, but whose exact evaluation
will require numerical work that will have to be much more advanced than
has been possible so far. In terms of this quantity the relevant analogue
of (\ref{16}) is expressible as 
\be {n\over\Theta^3}\approx \Big( {\sqrt{m_\ast}\over m}\Big)^3 
\big(\sqrt{m_\ast}\big)^{(3+\varepsilon)/(3-\varepsilon)}
\, ,\eqn{18}\fe
which means that the vorton number density $n$ will typically be very much
lower than in the preceeding case, but this is compensated, as far as their
contribution to the mass density of the universe
is concerned, by the fact that the typical vorton mass $M$ will be
much higher: the relevant analogue of (\ref{15}) is 
\be M\approx {m^{2}\over\ m_\ast} \, ,\eqn{19}\fe
which, in view of (\ref{17}), will always exceed the Planck mass.

Even if they contribute only a negligibly small fraction of the density of 
the universe, the vortons may nevertheless give rise to astrophysically
interesting effects: in particular it has recently been suggested by
Bonazzola and Peter\cite{BP97} that they might account for otherwise
inexplicable cosmic ray events.

\vfill\eject

\end{document}